\documentclass[aps,prl,superscriptaddress,twocolumn,a4paper,showpacs]{revtex4}
\usepackage{graphicx}
\usepackage{bm}

\newcommand{\dd}{\text{d}}

\begin{document}
\title{Measurement of the Neutrino Asymmetry Parameter $\bm{B}$ in Neutron Decay}
\date{\today}

\author{M.~Schumann}
\email{marc.schumann@gmx.net}
\affiliation{Physikalisches Institut der Universit\"at Heidelberg, Philosophenweg 12, 69120 Heidelberg, Germany}
\author{T.~Soldner}
\affiliation{Institut Laue-Langevin, B.P. 156, 38042 Grenoble Cedex 9, France}
\author{M.~Deissenroth}
\affiliation{Physikalisches Institut der Universit\"at Heidelberg, Philosophenweg 12, 69120 Heidelberg, Germany}
\author{F.~Gl\"{u}ck}
\affiliation{IKEP, Universit\"at Karlsruhe (TH), Kaiserstr.\ 12, 76131 Karlsruhe, Germany}
\affiliation{Research Institute for Nuclear and Particle Physics, POB 49, 1525 Budapest, Hungary}
\author{J.~Krempel}
\affiliation{Physikalisches Institut der Universit\"at Heidelberg, Philosophenweg 12, 69120 Heidelberg, Germany}
\affiliation{Institut Laue-Langevin, B.P. 156, 38042 Grenoble Cedex 9, France}
\author{M.~Kreuz}
\affiliation{Institut Laue-Langevin, B.P. 156, 38042 Grenoble Cedex 9, France}
\author{B.~M\"{a}rkisch}
\affiliation{Physikalisches Institut der Universit\"at Heidelberg, Philosophenweg 12, 69120 Heidelberg, Germany}
\author{D.~Mund}
\affiliation{Physikalisches Institut der Universit\"at Heidelberg, Philosophenweg 12, 69120 Heidelberg, Germany}
\author{A.~Petoukhov}
\affiliation{Institut Laue-Langevin, B.P. 156, 38042 Grenoble Cedex 9, France}
\author{H.~Abele}
\email{abele@physi.uni-heidelberg.de}
\affiliation{Physikalisches Institut der Universit\"at Heidelberg, Philosophenweg 12, 69120 Heidelberg, Germany}

\begin{abstract}
A new measurement of the neutrino asymmetry parameter $B$ in neutron decay, the
angular correlation between neutron spin and anti-neutrino momentum, is presented. The result,
\mbox{$B=0.9802(50)$}, confirms earlier measurements
but features considerably smaller corrections. It 
agrees with the Standard Model expectation and
permits updated tests on ``new physics'' in neutron decay.
\end{abstract}

\pacs{13.30.Ce; 12.60.Cn; 23.40.Bw; 24.80.+y}

\maketitle

Assuming only Lorentz invariance, 
the decay probability for polarized neutrons is given by \cite{Jac57}
\begin{eqnarray}
\label{jackson} \dd \omega &\propto& K \ \dd E  \ \dd \Omega_e \ \dd\Omega_\nu \nonumber 
 \left( 1 + a \ \frac{\bm{p}_e \bm{p}_\nu}{E E_\nu} + b \ \frac{m_e}{E} \right. \\
&& \left. + \langle \bm{s}_n
\rangle \left[ A \ \frac{\bm{p}_e}{E} + B \ \frac{\bm{p}_\nu}{E_\nu}+D \ \frac{\bm{p}_e \times
\bm{p}_\nu}{E E_\nu}\right]\right) \text{.}
\end{eqnarray}
$\bm{p}_e$, $\bm{p}_\nu$, $E$, $E_\nu$ are momentum and energy of electron and anti-neutrino (in the following called neutrino),
respectively, $m_e$ is the electron mass, $\langle \bm{s}_n \rangle$ is the neutron spin polarization, and the $\Omega_i$ denote solid angles.
The parameters $a$, $A$, $B$, and $D$ are angular correlation coefficients: 
$a$ is the correlation between the momenta of electron and neutrino. 
The parity violating parameters $A$ and $B$ for electron and neutrino
asymmetry correlate the neutron spin with the 
momentum of electron and neutrino, respectively. The subject of this paper is a precise measurement of $B$.

The factor $K$ and the correlation coefficients are related to the couplings
of the theory. Within the Standard Model of Particle Physics (SM), 
$b$ vanishes and $D$ is much smaller than the present experimental limit.
\mbox{$K=G_{\!F}^2 |V_{ud}|^2 F(E) (g_V^2 + 3 g_A^2)$}, with Fermi-constant $G_{\!F}$, quark
mixing matrix element $V_{ud}$, phase\-space factor $F(E)$, and vector and axial-vector coupling
constants $g_V$ and $g_A$. 
All correlation coefficients are functions of $\lambda\!=\!g_A/g_V$, where $g_A$ and $g_V$ are
assumed to be real, e.g.:
\begin{equation}\label{corr_coeff}
\quad A=-2\ \frac{\lambda^2 +\lambda}{1+3\lambda^2} \qquad \quad B=2 \ \frac{\lambda^2-\lambda}{1+3\lambda^2}\text{.}
\end{equation}
Small higher order corrections must be considered additionally.
The structure of the weak interaction ($V\!-\!A$ in the SM) is not
predicted by theory but has to be determined experimentally. Due to its sensitivity to the
neutrino helicity the neutrino asymmetry
parameter $B$ is an important input parameter for this purpose: Precise measurements 
of $B$ \cite{Kuz95,Ser98,Kre05b} can be used to derive limits on 
hypothetical right-handed current ($V\!+\!A$) contributions to $\beta$-decay \cite{Dub91,Ser93,Abe00} 
(mediated by new heavy charged bosons $W_R$).
These can be compared to 
other indirect measurements in muon \cite{Mus05} and nuclear $\beta$-decay 
\cite{Tho01} as well as to collider attempts to directly produce the $W_R$ 
\cite{Aba04} (cf.\ also \cite{PDG06}). When 
interpreted in general left-right symmetric (LRS) models beyond manifest LRS 
theory, information from all these experiments (direct, muon, and $\beta$-decay) are 
complementary and necessary to obtain limits \cite{Tho01}.

A deviation from eq.\ (\ref{corr_coeff}) is a signal for new physics not described
within the Standard Model and may be due to 
admixtures of right-handed currents, or due to anomalous (scalar, tensor) couplings
possibly caused by exotic models like leptoquark exchange \cite{Her01}. 
A recent review on this topic can be
found in \cite{Sev06}. Neutron decay, involving all particles of the first generation, 
is well suited to study the structure of the weak interaction \cite{Abe07} 
since theoretical corrections are small 
and well calculable as they do not depend on nuclear structure \cite{Sir67,Glu98,And04}. 

In order to measure the neutrino asymmetry parameter $B$, the
electron spectrometer PERKEO II \cite{Abe97}
was installed at the cold neutron beam position PF1B \cite{Abe06} at
the High Flux Reactor of the Institut Laue-Langevin (ILL), Grenoble. A cold neutron beam
with a thermal equivalent flux of $1.3\!\times\!10^{10}$ \mbox{n cm$^{-2}$ s$^{-1}$} 
was transversally spin polarized in a system of two supermirror polarizers in the new X-SM
geometry \cite{Kre05}. A radiofrequency (rf) spinflipper \cite{Baz93} allowed to reverse the spin direction.
Polarization $P$ and spinflipper \mbox{efficiency $F$}, formerly sources of large corrections and uncertainties,
were determined to $P=0.997(1)$ and $F=1.000(1)$ leading to a $0.30(14)\%$
correction on $B$. $P$ and $F$ were measured as a function of the neutron wavelength
by a time-of-flight method. A second rf-flipper and two supermirror analyzers
in the geometry of \cite{Kre05}
were used to determine $F$ and to scan $P$. For the precise measurement
of the absolute beam polarization, several opaque polarized $^3$He cells
\cite{Zim99,Zim99b} were employed. Both, $P$ and $F$ 
were uniform over the full neutron beam cross section for all wavelengths. 
The flipping ratio, a measure \mbox{for $PF$,} was monitored regularly and stayed constant during beamtime. 

Behind the polarizers, a neutron shutter was installed for 
background measurements in the electron detectors. It was made of enriched $^6$LiF,
as were the neutron beam collimation orifices and the beamstop at the end
of the installation, since generation of $\gamma$-radiation and fast neutrons 
in $^6$LiF is suppressed by $10^4$ \cite{Lon80}. Additional shielding was employed to
reduce the remaining background.

\begin{figure}
\includegraphics*[width=8.5cm]{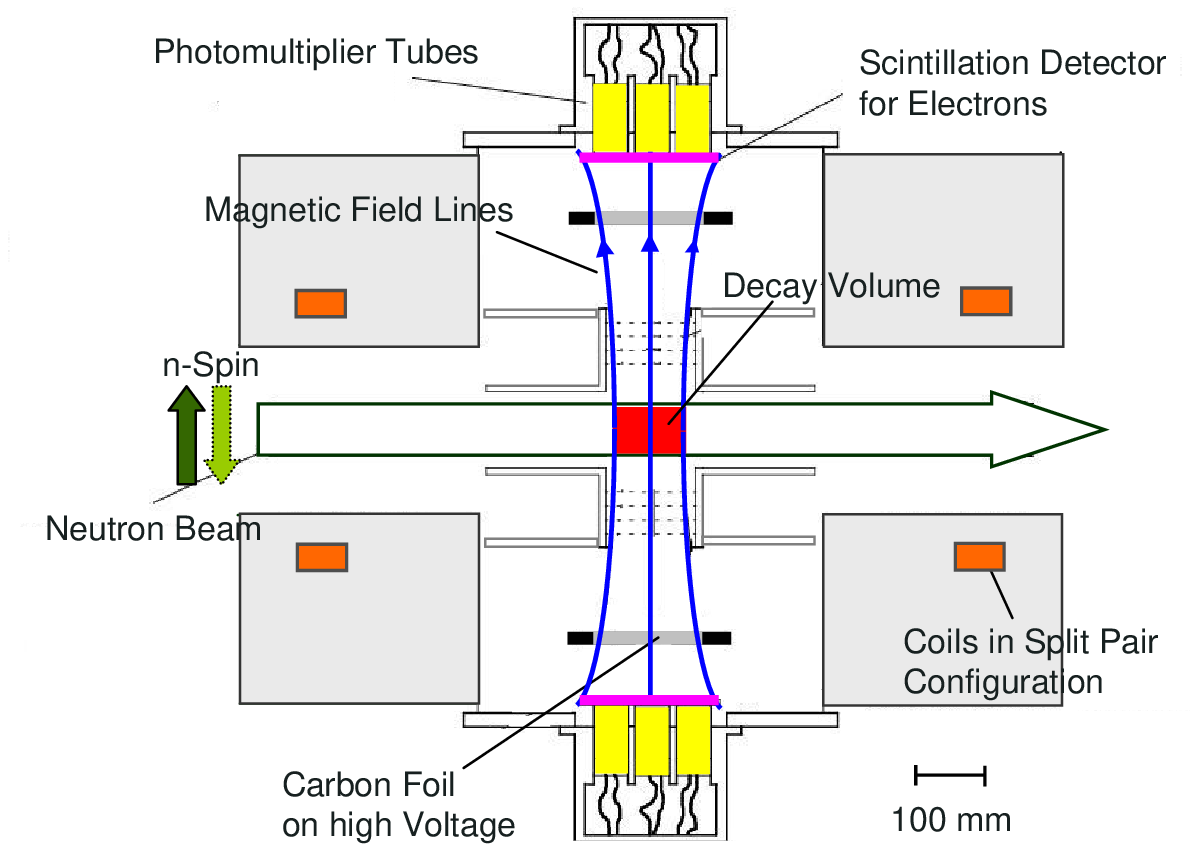} 
\caption{\label{Fig_Perkeo} The spectrometer PERKEO II: 
Polarized neutrons pass the setup, the magnetic field divides the full solid angle into two
hemispheres -- in and against spin direction -- and guides the decay products onto the detectors. 
The low energetic protons are accelerated onto a thin carbon foil on negative potential,
where they generate secondary electrons that can be detected by the scintillators.}  
\end{figure}

A certain fraction of neutrons decayed 
within the decay volume centered in the spectrometer.
Its main part is a pair of superconducting coils in split pair configuration,
generating a magnetic field with a maximum of 1.03 T perpendicular to the
neutron beam (fig.\ \ref{Fig_Perkeo}). The neutron spin aligns with the field that
therefore separates the full solid angle in two hemispheres: One in and one
against neutron spin direction. It guides the charged decay products onto the two detectors 
installed next to the beam, realizing a \mbox{2$\times$2$\pi$} detector where no solid angle corrections
have to be applied. Systematic effects related to the spectrometer design are described below. A detailed description of a previous experiment can be found in \cite{Kre05b}.

The magnetic field $B'$ decreases towards the detectors. This causes 
an increase of the parallel momentum
component of the particles leading to 
reduced electron backscattering. This is further reduced by the ``magnetic mirror effect'':
Electrons scattered out of the detector may be reflected at the increasing 
$B'$ and still
detected in the correct hemisphere. Backscattering 
is recognized via its delayed signal in the second 
detector. The full energy of the decay electron is reconstructed. The fraction 
of events assigned to the wrong detector is smaller than 0.2\%, a
neglectable systematic effect ($<10^{-4}$) 
if the region of interest is chosen above a $\beta$-energy of 240 keV 
\cite{Sch07}.

Since the neutrino cannot be measured directly, electron and proton
were detected in coincidence 
to reconstruct the neutrino. Most sensitive to $B$ is the case 
when electron and proton are emitted into the same hemisphere relative to the
neutron spin -- momentum conservation then restricts the neutrino to the
opposite direction. The other case, where electron and proton are emitted into 
different hemispheres, is kinematically favored but less sensitive to $B$
since the neutrino direction is less constrained \cite{Glu95}. As it depends strongly 
on detector calibration, this case was only 
used for cross checks of the result obtained for the first case.

Electrons ($E_{\text{max}}\!=\!782$ keV) are detected by 
\mbox{190 $\times$ 130 mm$^2$} plastic scintillators
with photomultiplier readout. The protons having much lower energies ($E^p_{\text{max}}\!=\!780$ eV) 
are accelerated onto a thin carbon foil (15$-$30 $\mu$g cm$^{-2}$, coated with MgO) 
on negative potential \mbox{($V\!=\!-18$ kV)}. Whereas
the electrons pass the foil almost unperturbed, the heavy protons have enough ionization
power to release one or several secondary electrons from the foil \cite{Kra66}. These
are detected in the scintillator, where also 
the proton time-of-flight is registered. 
No precise energy information on the proton is obtained with this method. 
Proton detection does not depend on the initial proton energy and the angles of
incidence occurring in this setup as was experimentally verified.

The measured signature is the experimental neutrino asymmetry defined by
\begin{equation}\label{expB}
B_{\text{exp}}(E) = \frac{N^{--}(E)-N^{++}(E)}{N^{--}(E)+N^{++}(E)}\text{,}
\end{equation}
where $N^{ij}(E)$ is the number of coincident events with elec\-tron kinetic energy $E$. The 
first/second sign 
indicates whether the electron/proton was emitted in ($+$) or against ($-$) neutron spin 
direction. Eq.\ (\ref{expB}) is related to
the neutrino asymmetry parameter $B$ by integrating eq.\ (\ref{jackson}) over the hemisphere \cite{Glu98,Glu95}
\begin{equation} \label{BFit}
B_{\text{exp}}(E) = \frac{4 P}{3} \left\{ \begin{array}{ll}
\frac{A \beta (2r\!-\!3)+B(3\!-\!r^2)}{8-4r+a \beta (r^2\!-\!2)} & \qquad  [r < 1] \\[0.1cm]
\frac{-A \beta +2 Br}{4r-a \beta} & \qquad  [r \ge 1].
\end{array} \right.
\end{equation} 
The definition is separated into two regions by the energy dependent parameter
$r\!=\!\beta (E\!+\!m_e)/(E_{\text{max}}\!-\!E)$
which is unity at $E=236$ keV. \mbox{$\beta=v/c$}. 
Eq.\ (\ref{BFit}) is very sensitive to the coefficient $B$ but also de\-pends 
slightly on the correlations $a$ and $A$ whose experimental uncertainties have to be considered.

Detector calibration was performed regularly and
two-dimensional detector scans were carried out to correct for spatial detector characteristics. 
Due to the flat spectral shape of $B_{\text{exp}}(E)$ detector calibration imposes
only a tiny uncertainty of 0.02\% on $B$.

At low electron energies, there is
background related to the high voltage (HV) applied on the carbon foils.
Above 230--240 keV, however, the measured electron spectra, i.e. $N^{++}(E)$ and 
$N^{--}(E)$, can be well described by their theoretical expressions,
where all fits have only one free parameter, a normalizing factor. 
An upper limit on remaining background contributions in the 
fit region was determined from the fit residuals.
At lower energies, a satisfactory description is impossible due to background, a non-linear 
energy calibration, and backscattered electrons assigned to the wrong detector \cite{Sch07b}.

All corrections due to the spectrometer design have been
obtained from Monte Carlo simulations. The ``edge effect'' accounts for the
loss of charged particles due to the finite length of the decay volume that was 
defined by thick aluminum baffles. ``Grid Effect'': 
Four layers of grids made of AlSi-wires 
(\mbox{10 $\mu$m} and 25 $\mu$m) were used to prevent the HV applied
to the detector foils to reach into the decay volume. Different methods 
(finite elements, boundary elements) showed that the absolute 
electric potential in the decay volume is at least one 
order of magnitude below a value that would cause systematic effects at
the present level of experimental precision. 
However, electrons and protons may be absorbed or scattered by the grids. This
``grid effect'' was obtained using the program PENELOPE \cite{Sem03} to simulate the 
electron trajectories in the wires. Protons hitting the wires 
were assumed to be absorbed. The potential
barrier for electrons to reach the scintillator can be neglected since 
all electrons with $E>84$ keV will certainly pass it regardless of their initial emission direction.

Charged particles moving in an increasing magnetic field $B'$ may be reflected as 
$p_\perp^2/B'$ is an adiabatic invariant, where $p_\perp$ is the momentum component perpendicular
to the field lines. This gives rise to the ``magnetic mirror effect'' since 
a certain fraction of decay products was emitted
towards the field maximum due to the finite neutron beam width. 
An asymmetric setup, i.e.\ a \mbox{displacement $\Delta$} 
between neutron beam and magnetic field maximum, may cause an additional, possibly large effect on $B$.
Therefore $\Delta$ 
was measured directly and was additionally determined from data in two
independent ways to correct for the effect: It was obtained from a $\chi^2$-minimization 
of fits to the difference spectrum $D(E)=N^{--}(E)-N^{++}(E)$ that has the highest sensitivity on $\Delta$ at
high electron energies $E$. $\Delta$ was also determined from 
the relative difference of the electron asymmetry parameters $A$ 
measured without ep-coincidence 
with the two detectors. This was possible due to the symmetric setup and since beam related
background, i.e. background generated in the collimation system that cannot be
measured separately, is small ($<10^{-4}$) and also symmetric 
in a region $E>350$ keV \cite{Mun06}. 
The resulting values for $\Delta$ do virtually not depend on $B$
nor on other systematic effects. The associated error of 0.32\% constitutes the
largest systematic uncertainty of the measurement.

\begin{table}[b]
\caption{\label{tab_bsame} Neutrino asymmetry $B$: corrections and errors} 
\begin{ruledtabular}
\begin{tabular}{l r r r r} 
 & \multicolumn{2}{c}{Detector 1} & \multicolumn{2}{c}{Detector 2}  \\
Effect [\%] & Corr. & Err. & Corr. & Err. \\ \hline
Polarization & +0.30 & 0.10 & +0.30 & 0.10 \\ \vspace{0.1cm} 
Flip Efficiency &  & 0.10 &  & 0.10 \\ 
Data Set: Statistics &  & 1.22 &  & 0.36 \\
\hspace{0.3cm}Proton Window & $-$0.05 & 0.03 & $-$0.05 & 0.03 \\
\hspace{0.3cm}1 Stop Condition &$-$0.24 & 0.06 & $-$0.13 & 0.03 \\ \vspace{0.1cm} 
\hspace{0.3cm}Background & & 0.10 & & 0.08 \\ \vspace{0.1cm} 
Detector Calibration &  & 0.02 &  & 0.02 \\ 
Spectrometer: Edge Effect & $-$0.16 & 0.05 & $-$0.16 &  0.05 \\
\hspace{0.3cm}Grid Effect & +0.03 & 0.05 & +0.03 & 0.05 \\
\hspace{0.3cm}Mirror Effect & +0.44 & 0.05 & +0.44 & 0.05 \\ \vspace{0.1cm} 
\hspace{0.3cm}Displacement $\Delta$ & $-$0.10 & 0.32 & +0.10 & 0.32 \\ 
Correlations $A$, $a$ & & 0.07 &  & 0.07 \\ \hline
Sum & $+$0.22 & 1.28 & $+$0.53 & 0.52 \\ 
\end{tabular}
\end{ruledtabular}
\end{table}

$F(E)$ in eq.\ (\ref{jackson}) has already been corrected for Coulomb 
interactions $F_C(E)$, proton recoil $R(E)$, and outer radiative corrections
$\delta_R(E)$, and reads
\begin{equation}
F(E)=F'(E) \ (1+\delta_R(E))  \ (1+R(E)) \ F_C(E) \text{,}
\end{equation} 
where $F'(E)$ is the uncorrected phasespace factor. The expressions for $\delta_R(E)$ and 
$R(E)$ were taken from \cite{Sir67} and \cite{Wil82}, respectively. 
The recoil and order-$\alpha$ corrections to $B$ are of order 0.01\% \cite{Glu98}.

The proton coincidence window $W_1$ was restricted to \mbox{40 $\mu$s}
causing a small correction of $-0.05(3)$\% to account
for slower protons. Accidental coincidences were 
directly measured with a delayed coincidence technique in a delayed window $W_2$
from 42$-$82 $\mu$s after the initial trigger. In order to avoid suppression of protons
by accidental coincidences or background, or suppression of accidental coincidences
by preceding signals, up to 32 stops were detected in both detectors.
Only events with exactly one stop in the respective window were considered 
in the analysis since multiple stops are mostly due to
background. Events with a stop in $W_1$ and a second 
(``accidental'') stop
in $W_2$ were included in the analysis.

However, this ``1 stop'' condition causes an overestimation of 
accidental coincidences since the stop-signal combination ``proton and accidental signal''
in $W_1$ 
is removed from the data set, whereas a similar combination does not occur in $W_2$. 
The available information on all stops allowed to determine the 
necessary correction directly from the data. 

\begin{figure}
\includegraphics*[width=8.5cm]{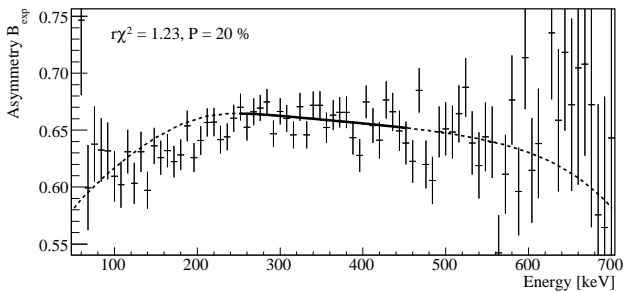}
\caption{\label{Fig_SameBD2} Fit of $B_{\text{exp}}$ to all detector 2 data. The solid curve
indicates the fit region. The result does not depend on this region. However, the overall
uncertainty increases if the fit is extended to higher energies due to the magnetic mirror
effect. }  
\end{figure}

At high electron energies $E$, the fit region is limited 
as the uncertainty related to the
displacement $\Delta$ between neutron beam and magnetic field increases
with $E$. At the low energy side, the fit region is
limited by the effects mentioned above: Background, non-linear detector response, and
wrongly assigned backscatter events. The region was chosen from 
250$-$455 keV. The final asymmetry parameter $B$ is independent of this choice
as the fit results agree within $\pm 0.3 \sigma_{\text{stat}}$ for intervals between 235 and 620 keV 
(fig.\ 4.48 in \cite{Sch07b}).

Fig.\ \ref{Fig_SameBD2} shows the fit of $B_{\text{exp}}$, eq.\ (\ref{BFit}), 
to all data of detector 2 (proton efficiency about 17\%). It yields the neutrino asymmetry parameter
\mbox{$B_2 = 0.9798(36)_{\text{stat}}(36)_{\text{syst}}$}. The result of
detector 1, 
\mbox{$B_1 = 0.9845(120)_{\text{stat}}(36)_{\text{syst}}$},
is limited by statistics due to a smaller proton
efficiency. This was caused by an inferior foil coating and higher HV background
that could not be further suppressed.
A detailed compilation of all relevant corrections and errors is given in table 
\ref{tab_bsame}.

In this situation, with two detectors of very different statistical significance, we use
the statistical average as the final neutrino asymmetry parameter result
\begin{equation}\label{eq_result}
B=0.9802(50) = 0.9802(34)_{\text{stat}}(36)_{\text{syst}} \text{.}
\end{equation}
The experiment is limited by statistics and the uncertainty due to the displacement $\Delta$
between neutron beam and magnetic field. 
With two detectors of equal performance, both errors would be significantly
smaller as the influence of $\Delta$ would cancel by calculating the
arithmetic mean of $B_1$ and $B_2$. 

The second case, electron and proton in opposite detectors, yields
results with much larger uncertainties, 1.9\% and 3.0\% for detector 1 and 2, respectively,
dominated by detector calibration. They statistically agree with (\ref{eq_result}).

Our result (\ref{eq_result}) has a precision similar to the most precise
measurement so far \cite{Ser98} and agrees very well with all results published earlier. 
It is distinguished, however, as it features several times smaller corrections than
competing experiments (0.5\%; 1\% if absolute numbers are considered, cf.\ table \ref{tab_bsame}). Our result is
consistent with the Standard Model expectation, \mbox{$B_{\text{SM}} = 0.9876(2)$}, 
calculated with the current world average for $\lambda$ from \cite{PDG06} and eq.\ (\ref{corr_coeff}).

By including our result (\ref{eq_result}), the error of the world average for $B$ reduces
by 25\%, yielding $\overline{B}=0.9807(30)$. We apply this value to analyze a manifest LRS
model with zero mixing ($\zeta=0$), following the procedure described in \cite{Abe00} but with
only the electron asymmetry parameter $A$ from \cite{PDG06} as further input parameter; $\lambda$
is a free parameter as it may be different from the SM value. We obtain a lower limit 
$m_{W_R}\!>\!290.7$ GeV/$c^2$ (90\% CL). The rather small improvement of this limit despite the
reduced uncertainty of $\overline{B}$ originates from the shift of $\overline{B}$ to a lower
value. Due to the controversial neutron lifetime (cf.\ \cite{PDG06}) we renounce a more elaborated
analysis. However, with this controversy being settled, the improved neutrino asymmetry parameter
together with other neutron decay data will permit to derive new limits on general LRS models and on
scalar and tensor couplings.

This work was funded by the German Federal Ministry for Research and Education, 
contract no. 06HD153I.



\begin{references}
\bibitem{Jac57} J. D. Jackson et al., Phys. Rev. {\bf106}, 517 (1957) 
\bibitem{Kuz95} I. A. Kuznetsov et al., Phys. Rev. Lett {\bf75}, 794 (1995)
\bibitem{Ser98} A. Serebrov et al., JETP {\bf86}, 1074 (1998) 
\bibitem{Kre05b} M. Kreuz et al., Phys. Lett. B {\bf619}, 263 (2005) 
\bibitem{Dub91} D. Dubbers, Prog. Part. Nucl. Phys. {\bf26}, 173 (1991) 
\bibitem{Ser93} A. Serebrov, N. Romanenko, JETP Lett. {\bf 55}, 503 (1992) 
\bibitem{Abe00} H. Abele, Nucl. Instr. Meth. A {\bf440}, 499 (2000) 
\bibitem{Mus05} J. R. Musser et al., Phys. Rev. Lett. {\bf94}, 101805 (2005) 
\bibitem{Tho01} E. Thomas et al., Nucl. Phys. A {\bf694}, 559 (2001) 
\bibitem{Aba04} V. M. Abazov et al., Phys. Rev. D {\bf69}, 111101(R) (2004) 
\bibitem{PDG06} W.-M. Yao et al. (PDG), J. Phys. G {\bf 33}, 1 (2006)
\bibitem{Her01} P. Herczeg, Prog. Part. Nucl. Phys {\bf 46}, 413 (2001) 
\bibitem{Sev06} N. Severijns, M. Beck, O. Naviliat-Cuncic, Rev. Mod. Phys. {\bf 78}, 991 (2006) 
\bibitem{Abe07} H. Abele, Part. Prog. Nucl. Phys.,\\ doi:10.1016/j.ppnp.2007.05.002 (2007) 
\bibitem{Sir67} A. Sirlin, Phys. Rev. {\bf 164}, 1767 (1967) 
\bibitem{Glu98} F. Gl\"uck, Phys. Lett. B {\bf 436}, 25 (1998)
\bibitem{And04} S. Ando et al., Phys. Lett. B {\bf595}, 250 (2004) 
\bibitem{Abe97} H. Abele et al., Phys. Lett. B {\bf 407}, 212 (1997) 
\bibitem{Abe06} H. Abele et al., Nucl. Instr. Meth. A {\bf562}, 407 (2006) 
\bibitem{Kre05} M. Kreuz et al., Nucl. Instr. Meth. A {\bf547}, 583 (2005) 
\bibitem{Baz93} A. N. Bazhenov et al., Nucl. Instr. Meth. {\bf332}, 534 (1993) 
\bibitem{Zim99} O. Zimmer et al., Phys. Lett B {\bf 455}, 62 (1999)
\bibitem{Zim99b} O. Zimmer, Phys. Lett. B {\bf 461}, 307 (1999) 
\bibitem{Lon80} M. A. Lone et al., Nucl. Instr. Meth. {\bf174}, 521 (1980) 
\bibitem{Sch07} M. Schumann, H. Abele, arXiv:0708.3150 (2007) 
\bibitem{Glu95} F. Gl\"uck, I. Jo\'{o}, J. Last, Nucl. Phys. A {\bf 593}, 125 (1995) 
\bibitem{Kra66} D. E. Kraus, F. A. White, IEEE Trans. Nucl. Sci. {\bf NS-13}, 765 (1966) 
\bibitem{Sch07b} M. Schumann, PhD thesis, University of Heidelberg 
	{\tt \small www.ub.uni-heidelberg.de/archiv/7357} (2007)
\bibitem{Sem03} J. Sempau et al., Nucl. Instr. Meth. B {\bf207}, 107 (2003) 
\bibitem{Mun06} D. Mund, PhD thesis, University of Heidelberg, 
	{\tt \small www.ub.uni-heidelberg.de/archiv/6576} (2006)
\bibitem{Wil82} D. H. Wilkinson, Nucl. Phys. A {\bf377}, 474 (1982)
\end{references}
\end{document}